\documentclass{iaus}
\usepackage{graphicx}

\title[Magnetism in very-low-mass stars] %% give here short title %%
{Magnetic field observations of low-mass stars}

\author[Reiners, A.]   %% give here short author list %%
{Ansgar Reiners$^1$}

\affiliation{$^1$ Universit\"at G\"ottingen, Institut f\"ur
  Astrophysik, Friedrich-Hund-Platz 1,\\ D-37077 G\"ottingen, Germany
  \break email: Ansgar.Reiners@phys.uni-goettingen.de}

\pubyear{2009}
\volume{259}  %% insert here IAU Symposium No.
\pagerange{119--126}
\date{?? and in revised form ??}
\jname{Cosmic Magnetic Fields: from Planets, to Stars and Galaxies}
\editors{K.G. Strassmeier, A.G. Kosovichev, \&, J. Beckmann, eds.}
\begin{document}

\maketitle

\begin{abstract}
  Direct measurements of magnetic fields in low-mass stars of spectral
  class M have become available during the last years. This
  contribution summarizes the data available on direct magnetic
  measurements in M dwarfs from Zeeman analysis in integrated and
  polarized light. Strong magnetic fields at kilo-Gauss strength are
  found throughout the whole M spectral range, and so far all field M
  dwarfs of spectral type M6 and later show strong magnetic fields.
  Zeeman Doppler images from polarized light find weaker fields, which
  may carry important information on magnetic field generation in
  partially and fully convective stars.

%% add here a maximum of 10 keywords, to be taken form the file <Keywords.txt>
\end{abstract}

\firstsection % if your document starts with a section,
              % remove some space above using this command.
\section{Introduction}

The existence and the topology of magnetic fields on low-mass stars is
a topic of great interest. On the Sun, we know that magnetic fields
are generated but strong fields are concentrated in small regions.
Many mechanisms that could generate these fields were suggested, but
their real physical nature still is a question of lively research (see
other contributions in this volume). Rotation is probably one of the
fundamental parameters that plays a key role in the generation of
magnetic fields (e.g., \cite{Pizzolato03}; \cite{Reiners09}). Other
paramaters that could matter are the luminosity of the stars
(\cite{Christensen09}), the convective velocities, the size of the
convection zone, and the existence of a region of great sheer between
a radiative (probably rigidly rotating) core and an outer convective
envelope, the so-called tachocline (\cite{Ossendrijver03}).

The general paradigm of the solar dynamo assumes that at least the
cyclic part of the solar magnetic field is generated close to the
tachocline through a so-called $\alpha\Omega$-type dynamo (but see
also Brandenburg, this volume). Any dynamo mechanism that requires a
tachocline must vanish in low-mass stars around spectral type M3
because this is the boundary beyond which stars no longer maintain a
radiative core but remain fully convective during their entire
lifetime (very young age, more massive stars are also fully
convective). It is therefore particularly interesting to investigate
magnetic fields around the mass boundary between partially and fully
convective stars.

\section{Methods to measure magnetic fields}

Magnetic fields are measured with different techniques. Here, we
consider only ``direct'' measurements of magnetic fields, i.e., we are
not considering magnetic fields that are inferred through the
observation of secondary indicators. Such indicators are very useful
because they trace non-thermal radiation processes, which are probably
connected to the existence of magnetic fields (most prominent tracers
are chromospheric Ca~H\&K and H$\alpha$ emission, e.g.,
\cite{Hartmann87, Mohanty03}; coronal X-ray emission, e.g.,
\cite{Pizzolato03}; or radio emission from high-energy electrons,
\cite{Berger06, Hallinan08}).

Direct observations of magnetic fields can be accomplished through
observation of the Zeeman effect, which shows direct consequences of
magnetic fields on the appearance of spectral features. Principally,
two different ways must be distinguished. The first method is the
measurement of the Zeeman effect in polarized light, usually detecting
circular polarization in Stokes~V.  This method can detect relatively
weak fields because it is a differential method that can be calibrated
rather accurately. One caveat is that only a net polarization can be
detected. Polarization signals even of strong magnetic fields can
cancel out each other if they are distributed in small entities.
Another problem of measurements in Stokes~V is that only a fraction of
the light can be used so that the usually a number of simplifications
have to be applied. The second method is to determine the magnetic
fields strength from the appearance of spectral lines in Stokes~I,
i.e., from integrated light without any polarization analysis. Here,
the main problem is that the effect of Zeeman broadening is relatively
small in comparison to other broadening effects in spectral lines so
that calibration is very difficult and leads to large uncertainties.
On the other hand, this method can detect the total mean unsigned
magnetic flux including components that are distributed in small cells
of opposite polarity.

\section{A compilation of results}

\begin{table}
%\centering
\begin{minipage}[t]{\columnwidth}
  \caption{\label{tab:Mfields}Stokes I measurements of magnetic flux
    among M dwarfs.}
  \centering
  \begin{tabular}{lcrrr}
    \noalign{\smallskip}
    \hline
    \hline
    \noalign{\smallskip}
    Name & Spectral Type & $v\,\sin{i}$  & $Bf$ & Ref\\
         &               &[km\,s$^{-1}$] & [kG]\\
    \hline
    \noalign{\smallskip}   
    Gl~182      & M0.5 &    9       &     $ 2.5$ & $^{e}$ \\
    Gl~494A     & M0.5 &   11       &     $ 3.0$ & $^{e,f}$ \\
    Gl~70       & M2.0 & $\le$3     &     $ 0.0$ & $^{a}$\\
    Gl~569A     & M2.5 & $\le$3     &     $ 1.8$ & $^{e}$ \\
    Gl~729      & M3.5 &    4       &     $ 2.2$ & $^{a,b}$ \\
    Gl~873      & M3.5 & $\le$3     &     $ 3.9$ & $^{a,b}$ \\
    Gl~388      & M3.5 & $\approx$3 &     $ 2.9$ & $^{a,b}$ \\
    GJ~3379     & M3.5 & $\le$3     &     $ 2.3$ & $^{c}$ \\
    Gl~876      & M4.0 & $\le$3     &     $ 0.0$ & $^{a}$ \\
    GJ~1005A    & M4.0 & $\le$3     &     $ 0.0$ & $^{a}$ \\
    GJ~2069B    & M4.0 &    6       &     $ 2.7$ & $^{c}$ \\
    GJ~299      & M4.5 & $\le$3     &     $ 0.5$ & $^{a}$ \\
    GJ~1227     & M4.5 & $\le$3     &     $ 0.0$ & $^{a}$ \\
    GJ~1224     & M4.5 & $\le$3     &     $ 2.7$ & $^{a}$ \\
    Gl~285      & M4.5 &    5       &     $>3.9$ & $^{a,b}$ \\
    Gl~493.1    & M4.5 &   18       &     $ 2.1$ & $^{c}$ \\
    LHS~3376    & M4.5 &   19       &     $ 2.0$ & $^{c}$ \\
    Gl~905      & M5.0 & $\le$3     &     $ 0.0$ & $^{a}$ \\
    GJ~1057     & M5.0 & $\le$3     &     $ 0.0$ & $^{a}$ \\
    GJ~1154A    & M5.0 &    6       &     $ 2.1$ & $^{c}$ \\
    GJ~1156B    & M5.0 &   17       &     $ 2.1$ & $^{c}$ \\
    LHS~1070 A  & M5.5 &    8       &     $ 2.0$ & $^{c}$ \\
    GJ~1245B    & M5.5 &    7       &     $ 1.7$ & $^{a}$ \\
    GJ~1286     & M5.5 & $\le$3     &     $ 0.4$ & $^{a}$ \\
    GJ~1002     & M5.5 & $\le$3     &     $ 0.0$ & $^{a}$ \\
    Gl~406      & M5.5 &    3       &     2.1--2.4 & $^{a,d}$ \\
    GJ~1111     & M6.0 &   13       &     $ 1.7$ & $^{a}$ \\
    GJ~412B     & M6.0 &    5       &     $>3.9$ & $^{c}$ \\
    VB~8        & M7.0 &    5       &     $ 2.3$ & $^{a}$ \\
    LHS~3003    & M7.0 &    6       &     $ 1.5$ & $^{a}$ \\
    LHS~2645    & M7.5 &    8       &     $ 2.1$ & $^{a}$ \\
    LP~412$-$31 & M8.0 &    9       &     $>3.9$ & $^{a}$ \\
    VB~10       & M8.0 &    6       &     $ 1.3$ & $^{a}$ \\
    LHS~1070 B  & M8.5 &   16       &     $ 4.0$ & $^{c}$ \\
    LHS~1070 C  & M9.0 &   16       &     $ 2.0$ & $^{c}$ \\
    LHS~2924    & M9.0 &   10       &     $ 1.6$ & $^{a}$ \\
    LHS~2065    & M9.0 &   12       &     $>3.9$ & $^{a}$ \\
    \noalign{\smallskip}
    \hline
    \noalign{\smallskip}
  \end{tabular}
\end{minipage}
$^a$\cite{RB07},
$^b$\cite{JKV00},
$^c$\cite{Reiners09},
$^d$\cite{Reiners07},
$^e$\cite{RB09},
$^f$\cite{Saar96}
\end{table}

This article gives an overview of recent (direct) magnetic field
measurements in low-mass stars of spectral class M. Today, the number
of measurements is constantly growing and we can start to compare
results from different techniques in the same stars. This opens
another channel of investigation because the differences between the
results from different techniques carries information about the
magnetic fields and their distribution.
 
\subsection{Stokes~I}

Measurements of stellar magnetic fields in integrated light, Stokes~I,
can be carried out by calculating the polarized radiative transfer in
magnetically sensitive spectral lines and comparing the models to
observations. \cite{Robinson80}, has introduced the method to measure
stellar magnetic fields in spectral absorption lines. This method was
used by several groups, mostly in atomic spectral lines in the optical
wavelength range. A summary of this effort can be found in
\cite{Saar96}, and \cite{Saar01}. \cite{JKV00}, summarize results on
magnetic field measurements including spectral lines in the infrared
wavelength range. These authors center on developments for fully
convective M dwarfs and pre-main sequence stars. In very cool
atmospheres, e.g., in mid-M dwarfs at the boundary to full convection,
the usefulness of atomic lines becomes limited because atomic lines
are being buried under the haze of molecular absorption bands.

\cite{Valenti01}, discuss molecular absorption bands of FeH as a
Zeeman diagnostic for very cool dwarfs. FeH shows a series of narrow,
strong absorption lines for example around 1\,$\mu$m, a spectral
region where the spectral energy distribution of M dwarfs is close to
its maximum.  A problem for magnetic field measurements with FeH is
that magnetic splitting is difficult to calculate. \cite{RB06},
introduced a semi-empirical approach in which the magnetic field of an
M-star is measured by comparison of its FeH lines to FeH lines
observed in stars of known magnetic field. The magnetic fields of
these template stars have to be measured in atomic lines, hence only
early-M stars can be used for this method. \cite{RB06}, show that FeH
absorption in M dwarfs follows an optical-depth scaling that allows to
use early-M template spectra to measure magnetic fields even in
very-late type M dwarfs. Using this method, \cite{RB07}, carried out
the first survey of magnetic field measurements in M dwarfs, and
\cite{Reiners07}, and \cite{RB09}, applied the same method to another
set of stars. The uncertainty of their magnetic field measurements is
on the order of a kilo-Gauss so that small fields of a few hundred
Gauss are difficult to detect. The method was successfull in detecting
strong (several kilo-Gauss) fields in many M dwarfs, in particular all
M dwarfs with spectral type M6 and later show strong magnetic fields.

The current sample of M dwarfs with measured magnetic fields is
collected in Table\,\ref{tab:Mfields}. Strong magnetic fields are
found throughout the whole range of spectral subclasses M0--M9. More
specific, among the early-M stars with spectral type M0--M5.5,
magnetic flux is found between 0 and 4\,kG. Magnetic flux scales with
rotation (and activity in H$\alpha$ scales with magnetic flux), which
is expected because of the well-known rotation-activity connection
among warmer stars (e.g., \cite{Pizzolato03}). The meaning of this is
threefold: 1) M dwarfs (partially as well as fully convective) can
generate strong magnetic fields; 2) Stellar activity is of magnetic
origin; 3) The connection between rotation and magnetic field
generation / magnetic activity also holds in early- to mid-M dwarfs.

Magnetic field measurements have also been carried out in young
low-mass stars (see for example \cite{JK07}). Measurements of magnetic
fields in young objects are the subject of \cite{JK09}, in this
volume.

\subsection{Stokes~V}

The method of Zeeman Doppler Imaging in Stokes~V has been very
successfull during the last years providing the first information on
stellar magnetic structure (e.g., \cite{Donati97}). The somewhat
surprising result of axisymmetric magnetic topology in a fully
convective star was reported by \cite{Donati06} using this technique.
Recently, \cite{Donati08}, and \cite{Morin08}, measured magnetic
fields in Stokes~V in a sample of early- and mid-M dwarfs on both
sides of the boundary to full convection. They find magnetic flux on
the level of a few hundred Gauss in their sample stars. An important
result of this work is that the strength of magnetic flux and the
topology of magnetic fields show a marked change at spectral class M3
where stars are believed to become fully convective. This is very
interesting because it was expected that magnetic fields in fully
convective stars (\cite{Durney93}) are less organized than in stars
with a radiative core that might generate large-scale dipolar fields
through an interface dynamo located at the tachocline
(\cite{Charbonneau05}; \cite{Ossendrijver03}).

\section{A comparison of results}

\begin{figure}
 \resizebox{.5\hsize}{!}{\includegraphics{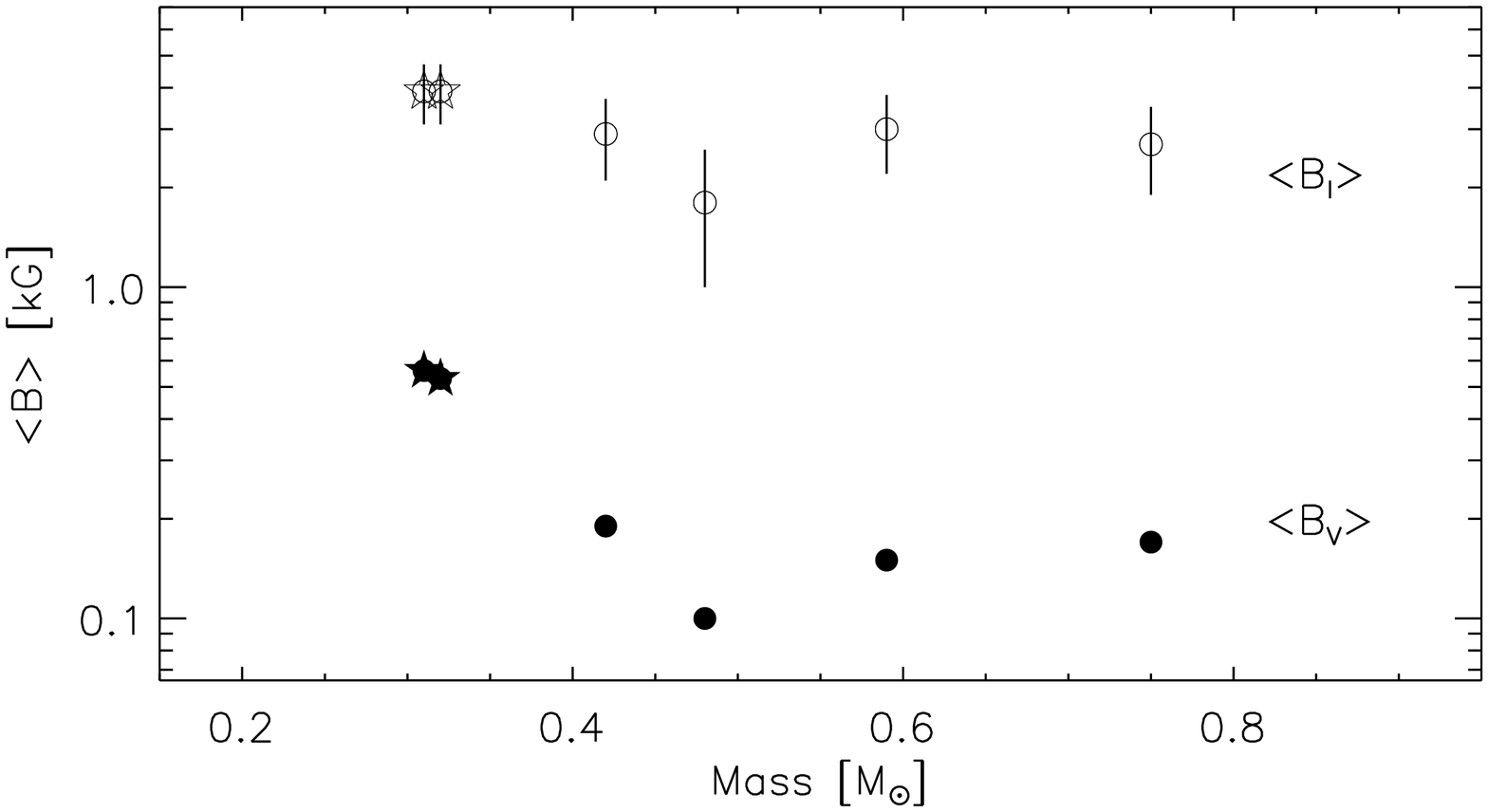}}
 \resizebox{.5\hsize}{!}{\includegraphics{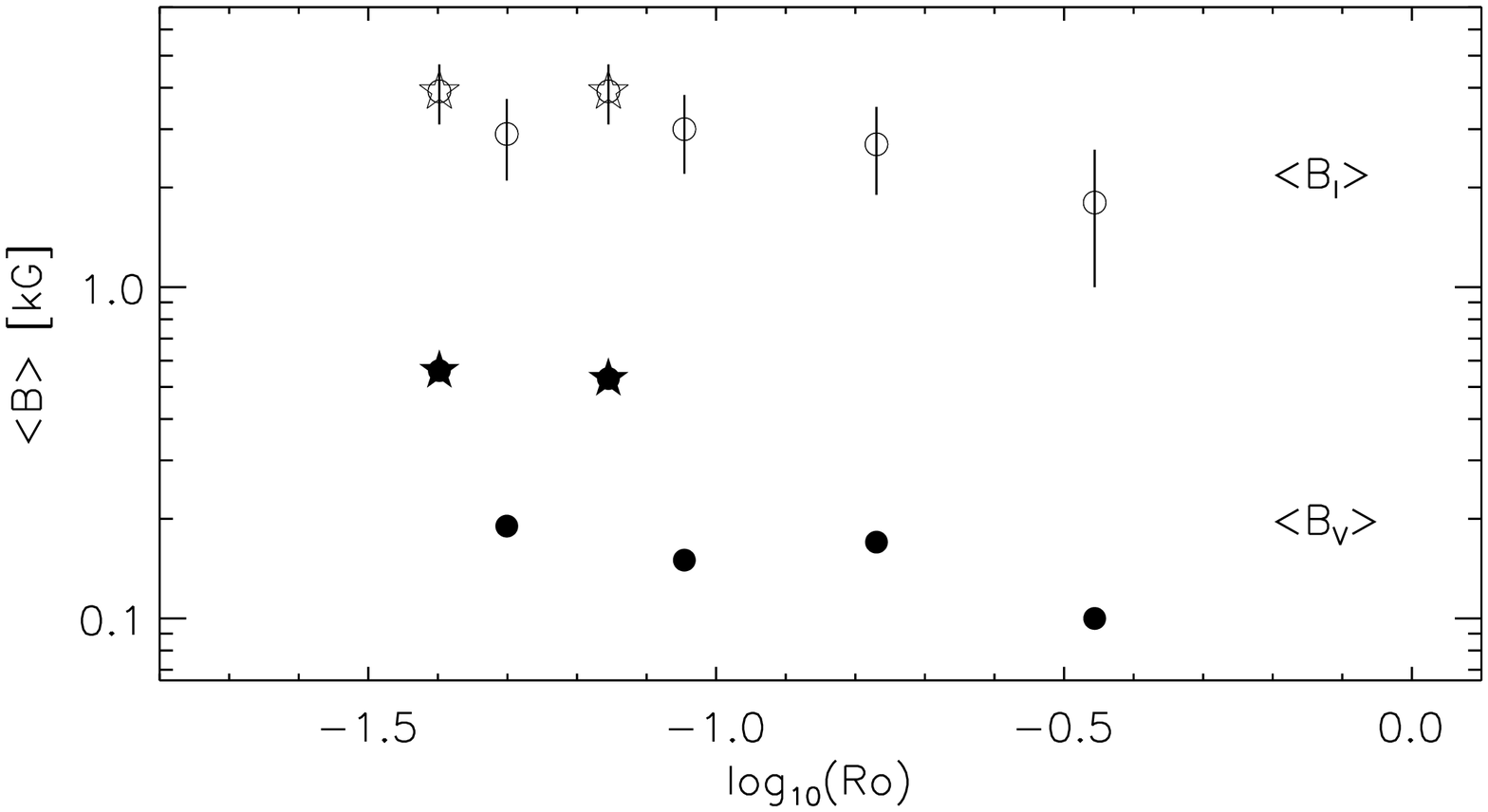}}
 \caption{\label{fig:comparison}Mean magnetic field measurements from
   Stokes~I (open symbols) and Stokes~V (filled symbols) as a function
   of Mass (left panel) and Rossby number (right panel).}
\end{figure}

With the growing number of magnetic field measurements in both,
integrated light and Zeeman Doppler Imaging from Stokes~V, it becomes
possible to compare results from both techniques. This may shed more
light on the topology of M star magnetic fields because both
techniques are sensitive to different aspects of the flux: While in
Stokes~I the mean total unsigned flux is observed, Stokes~V is
sensitive to the net polarized flux, i.e., to magnetic field
distributions that lead to a net polarization signal in time series
taken for the Doppler Images.

Currently, six M dwarfs have magnetic flux measurements from both
Stokes~V and Stokes~I. These results are shown in
Fig.\,\ref{fig:comparison} as a function of mass and Rossby number. In
all cases, the magnetic flux measured from Stokes~I is much stronger
than the value from Zeeman Doppler Imaging. This is not surprising, as
mentioned above, because the former is sensitive to the total flux
while the latter normally is not. The ratio between magnetic flux in
Stokes~V and I is between 1:7 and 1:17, this means the fraction of
magnetic flux seen in Stokes~V is between 6\,\% and 15\,\%.

\section{Summary}

Magnetic field measurements are becoming available for a growing
number of low-mass stars through analysis of integrated and polarized
light. A compilation of results from Stokes~I measurements in the FeH
molecule used among M dwarfs is given in this article. Other
techniques are available probing different aspecs of the magnetic
field distribution and making different assumptions. 

Generally, fields of kilo-Gauss strength are ubiquitously found in
late-M dwars, which is consistent with their activity. Magnetic field
topology is important to answer the question about the underlying
dynamo mechanism, and first results on topologies are available from
Zeeman Doppler imaging finding predominantly axisymmetric geometries
in fully convective stars.  However, the applied techniques are
probably missing a substantial fraction of the total magnetic flux so
that the full distribution of the magnetic fields may not be visible.

Brown dwarfs are generally rapidly rotating and so far no useful
tracer for Zeeman analysis was found (at least in old brown dwarfs),
which poses a serious problems to the detection of magnetism.  Whether
brown dwarfs can maintain strong magnetic fields is an open question,
but so far there seems to be no reason to believe they wouldn't.

\begin{acknowledgments}
  The author acknowledges research funding through a DFG Emmy Noether
  Fellowship (RE~1664/4-1).
\end{acknowledgments}

\end{document}